\documentclass[12pt,psfig,epsfig,graphicx,psfrag]{article}
\usepackage{dcolumn}
\usepackage{amssymb}
\usepackage{lscape}
\usepackage{amsmath}

\newcommand{\ba}{\begin{eqnarray}}
\newcommand{\ea}{\end{eqnarray}}
\newcommand{\be}{\begin{equation}}
\newcommand{\ee}{\end{equation}}
\newcommand{\thetavierbein}{E}
\newcommand{\fvierbein}{L}
\newcommand{\fmoinsunvierbein}{\ell}
\newcommand{\E}{F}
\newcommand{\cc}{c}
\newcommand{\w}{w}
\newcommand{\wb}{\bar{w}}
\newcommand{\W}{W}

\title{Covariant constraints in ghost free massive gravity}

\begin{document}
\maketitle

\begin{center}

 \vspace{0.5cm}
 {\large
C.~Deffayet$^{a,}$\footnote{deffayet@iap.fr}, J.~Mourad$^{a,}$\footnote{mourad@apc.univ-paris7.fr}, G.~Zahariade$^{a,}$\footnote{zahariad@apc.univ-paris7.fr}\\}
$^a${\it APC\;\footnote{UMR 7164 (CNRS, Universit\'e Paris 7, CEA, Observatoire de Paris)}, 10 rue Alice Domon et L\'eonie Duquet,\\}
 75205 Paris Cedex 13, France.\\
\date{\today}
\vspace{0.5cm}
 \bigskip

{\bf \large Abstract}
\begin{quotation}\noindent
 We show that the reformulation of the de Rham-Gabadadze-Tolley massive gravity theory using vielbeins leads to a very simple and covariant way to count constraints, and hence degrees of freedom. Our method singles out a
subset of theories, in the de Rham-Gabadadze-Tolley family, where an extra
constraint, needed to eliminate the Boulware Deser ghost, is easily seen
to appear. As a side result, we also introduce a new method, different from the Stuckelberg trick, to extract kinetic terms for the polarizations propagating in addition to those of the massless graviton.
\end{quotation}

\end{center}

\maketitle

\section{Introduction}
During the past few years and in particular following discussions of the DGP model \cite{Dvali:2000hr} and its cosmology \cite{Deffayet:2000uy,Deffayet:2001pu}, there has been renewed interest in theories of ``massive gravity" 
(see e.g. \cite{Rubakov:2008nh,Hinterbichler:2011tt} for reviews). The unique consistent theory for a free massive spin-2 field was known for a long time to be the Fierz-Pauli theory \cite{Fierz:1939ix}. This theory propagates 5 degrees of freedom of positive energy, one of which is a zero helicity polarization responsible for the celebrated van-Dam Veltman Zakharov discontinuity: namely that, however small the graviton mass, Fierz-Pauli theory leads to different physical predictions (such as light bending) from those of linearized General Relativity \cite{vanDam:1970vg}. Considering  self-interactions of the massive graviton leads to a mechanism, first discussed by Vainshtein \cite{Vainshtein:1972sx}, which can actually restore the continuity towards well tested predictions of General Relativity \cite{Deffayet:2001uk,Babichev:2009us,Babichev:2009jt,Babichev:2010jd}. However, massive graviton self-interactions also introduce a new pathology: the fact that a ghost-like 6th degree of freedom propagates generically in the full non linear theory, as was first pointed out by Boulware and Deser \cite{Boulware:1973my}. It was long thought impossible to obtain a massive gravity theory devoid of this ghost (see e.g. \cite{Boulware:1973my,Creminelli:2005qk}). However, a family of massive gravity theories was recently proposed by de Rham Gabadadze and Tolley (dRGT in the following) \cite{deRham:2010kj,deRham:2010ik,deRham:2011rn} in which the absence of ghost was first pointed out in the so-called decoupling limit \cite{deRham:2010ik} (using, in particular, the approach of Refs. \cite{Creminelli:2005qk,ArkaniHamed:2002sp,Deffayet:2005ys}) and then fully confirmed at the nonlinear level \footnote{Note, however, that these results have been debated \cite{Alberte:2010qb,Chamseddine:2011mu}.} by a Hamiltonian analysis, later extended to bimetric theories \cite{Hassan:2011ea,Hassan:2011zd,Hassan:2011tf,Hassan:2011hr} (see also \cite{Kluson,OthersCounting}).  The Hamiltonian analysis of these models remains however complicated and does not clarify the reasons behind their soundness. 

Here, we will show that the reformulation of dRGT models in terms of vierbeins leads to a simple way of extracting covariant (Lagrangian) constraints. Some of the Hamiltonian properties of such a reformulation have already been analyzed in \cite{Hinterbichler:2012cn} \footnote{See also \cite{Hassan:2012wt}.} (building on the older work of \cite{Nibbelink:2006sz}), where it was underlined in particular that one of the two necessary supplementary Hamiltonian constraints was easier to obtain with vierbeins than with the metric formulation. The analysis presented here is, however, different from the one presented in Ref. \cite{Hinterbichler:2012cn}, in particular because it is fully Lagrangian and also because some of the arguments given there are completed. 

This paper is organized as follows. We first recall (section \ref{S1}) how one can count degrees of freedom of massive gravity using covariant constraints. We then (section \ref{S2}) introduce the dRGT theories both in the metric and in the vierbein formulation. We  then present our covariant way of obtaining constraints (section \ref{S4}), which is closely related to what can be done in quadratic Fierz-Pauli theory.  As a side  result, we also introduce a new trick, different from Stuckelberg's, to extract kinetic terms for the polarizations propagating in addition to those of the massless graviton. This trick, valid for a subset of vierbein formulated dRGT theories, is presented in appendix \ref{appC}.

\section{The Boulware-Deser ghost from covariant constraints counting.}
\label{S1}

Let us first introduce the (quadratic) Fierz-Pauli theory \cite{Fierz:1939ix}. This can be defined on a flat space-time \footnote{For simplicity, we only discuss in this section the case with $D=4$ dimensions, but such a theory can be introduced in a similar way for arbitrary $D$, where it is found that a massive graviton in $D$ dimensions, as described by the quadratic Fierz-Pauli theory, has $(D-2)(D+1)/2$ physical polarizations, i.e. the same number of polarizations as a massless graviton in $D+1$ dimensions.} by the following action (in the absence of matter source) for a rank-2 covariant tensor $h_{\mu \nu}$ 
\ba \label{PF}
S_{h,m} & =& - M_h^2 \int d^4x \left[ \left(\partial_\mu h_{\nu\rho} \right)^2 - \left(\partial_\mu h\right)^2 \nonumber
+ 2 \left(\partial_\mu h\right)\left( \partial^\nu h^\mu_\nu \right)- 2 \left(\partial_\mu h_{\nu \rho}\right) \left(\partial^\nu h^{\mu \rho}\right)\right.\nonumber \\
&&\left. + m^2\left( h_{\mu \nu}h^{\mu \nu}-h^2\right)\right].
\ea 
Here $M_h$ is a mass parameter, all indices are moved up and down with a flat canonical Minkowski metric $\eta_{\mu \nu}$, and $h$ is defined by $h \equiv h_{\mu \nu} \eta^{\mu \nu}$. The terms on the right hand side of the first line of Eq. (\ref{PF}) are obtained by expanding the Einstein-Hilbert Lagrangian density $\sqrt{-g} R(g)$ at quadratic order around a flat metric, writing $g_{\mu \nu} = \eta_{\mu \nu} + h_{\mu \nu}$.
The mass terms appear in the second line on the right hand side of Eq. (\ref{PF}) and this particular combination of $h^2$ and $(h_{\mu \nu})^2$ is the only one able to give a mass to the graviton $h_{\mu \nu}$ in a consistent and ghost-free way.  Note that this theory explicitly breaks general covariance and also that it uses two rank-2 covariant tensors, $h_{\mu \nu}$ as well as $\eta_{\mu \nu}$ which serves as a background on which $h_{\mu \nu}$ propagates. 

Fierz-Pauli theory can be non linearly completed by considering actions of the form 
\ba \label{NLPFG} 
S_{g,m} = M_g^2 \int d^4 x \sqrt{-g} \left[R(g) -  m^2 V\left({\cal M}\right)\right],
\ea
where $V$ is a suitably chosen scalar function of ${\cal M}^\mu_{\hphantom{\mu} \nu} = g^{\mu \sigma} f_{\sigma \nu}$, $m$ and $M_g$ are again mass parameters, $R(g)$ is the Ricci scalar constructed from the metric $g_{\mu \nu}$, and the theory contains, besides the dynamical metric $g_{\mu \nu}$, a non dynamical metric $f_{\mu \nu}$ usually considered to be flat. If one wants to consider a (non linear) ``massive gravity" the potential $V$ should be chosen such that (i) when $f_{\mu \nu}$ is taken to be $\eta_{\mu \nu}$, $g_{\mu \nu} = \eta_{\mu \nu}$ is a solution of the field equations, and (ii) when expanded at quadratic order around this flat background, the action (\ref{NLPFG}) has the Fierz-Pauli form (\ref{PF}).  Note that when one has two metrics, one can write any non trivial non derivative invariant built from the metrics as a function of ${\cal M}$, and hence the only restriction (besides diffeomorphism invariance) comes here from requirements (i) and (ii) on $V$. Note also that it is easy to figure out that there are infinitely many functions $V$ that satisfy these requirements (see e.g. \cite{Damour:2002ws}).

As first noticed by Boulware and Deser \cite{Boulware:1973my}, quadratic Fierz-Pauli theory (\ref{PF}) and its non linear version (\ref{NLPFG}) differ dramatically as far as the number of propagating degrees of freedom is concerned. Consider first the former theory. Varying action (\ref{PF}) with respect to $h_{\mu \nu}$ one easily obtains the field equations  
\ba \label{FIELDgen}  \partial_\mu \partial_\nu h + \Box h_{\mu \nu} - \partial_\rho \partial_\mu h^\rho_\nu -\partial_\rho \partial_\nu h^\rho_\mu  + \eta_{\mu \nu}(\partial^\rho \partial^\sigma h_{\rho \sigma} - \Box h)  =  m^2 \left(h_{\mu \nu} - h \eta _{\mu \nu}\right).
\ea
Here the left hand side is just the linearized Einstein tensor and hence, as a consequence of Bianchi identities, its divergence $\partial^\mu$ vanishes. Thus, from the right hand side of Eq.  (\ref{FIELDgen}), we get
\ba \label{EQAUX1}
 \partial^\mu h_{\mu \nu} - \partial_\nu h = 0.
\ea
Taking another derivative of this equation yields
\ba \label{EQAUX2}
\partial^\nu \partial^\mu h_{\mu \nu} - \Box h = 0,
\ea 
where the combination appearing in the left hand side is just the linearization of the Ricci scalar. Thus contracting both sides of Eq. (\ref{FIELDgen}) with $\eta^{\mu \nu}$ yields   
\ba \label{traceEQ}
\Box h - \partial^\nu \partial^\mu h_{\mu \nu} = \frac{3}{2} m^2 h,
\ea
which together with (\ref{EQAUX2}) shows that $h$ vanishes in vacuum. This in turn means, using (\ref{EQAUX1}) that $h_{\mu \nu}$ is transverse. 
The two equations we just obtained, namely, 
\ba
\partial^\mu h_{\mu \nu} &=&  0, \label{CONS1}\\
h&=& 0   \label{CONS2},
\ea
together give $5$ Lagrangian constraints (being first order)  and this removes $5$ of the {\it a priori} $10$ dynamical degrees of freedom of $h_{\mu \nu}$. Hence, 
quadratic Fierz-Pauli theory propagates 5 polarizations. A similar conclusion can also be reached using a Hamiltonian counting (see e.g. \cite{Boulware:1973my}). 

In contrast, a generic non linear massive gravity propagates in addition a sixth ghost-like polarization. This was first argued by Boulware and Deser in Ref. \cite{Boulware:1973my} and thus the extra propagating mode is usually called a Boulware-Deser ghost. 
Schematically, this comes from the fact that the analog of the constraint (\ref{CONS2}) is lost, while there are still $4$ constraints similar to (\ref{CONS1}).   Indeed, now vary action (\ref{NLPFG}) with respect to $g_{\mu \nu}$ to obtain, 
\ba \label{EOM}
G_{\mu \nu} = m^2 T_{\mu \nu}^{(g,f)}
\ea
where $G_{\mu \nu}$ is the Einstein tensor built from the metric $g$, and $T_{\mu \nu}^{(g,f)}$ is obtained from varying the term coupling the two metrics $\sqrt{-g} V({\cal M})$, so that it contains no derivatives. Taking the covariant derivative $\nabla^\mu$ (with respect to the metric $g_{\mu\nu}$) of both sides of the above equation, gives 
\ba \label{CONSNL}
\nabla^{\mu} T_{\mu \nu}^{(g,f)} = 0 
\ea
which therefore only contains first derivatives\footnote{Note that the derivatives of $g$ only appear in the Christoffel symbols of $g$ whenever a $g$ covariant derivative hits an $f_{\mu \nu}$ metric, since the action of such a derivative on $g_{\mu \nu}$ vanishes, and hence yields no derivatives.} (and in particular no derivatives of the metric $f_{\mu\nu}$ if this metric is just taken to have the canonical Minkowski form $\eta_{\mu \nu}$) and hence yields 4 constraints on the dynamical metric $g_{\mu \nu}$. These constraints are the analog of (\ref{CONS1}) in the non linear case. On tracing over Eq.(\ref{EOM}) and using derivatives of Eq.(\ref{CONSNL}), there is now (as opposed to linear Fierz-Pauli theory) no reason to get an extra constraint (cf. also \cite{Deffayet:2005ys}). And in fact, it was thought impossible to construct a non linear Fierz-Pauli theory, with a suitable potential $V$, devoid of the Boulware Deser mode \cite{Boulware:1973my,Creminelli:2005qk} until the work of de Rham, Gabadadze and Tolley (henceforth dRGT) \cite{deRham:2010kj,deRham:2010ik}.

\section{The metric and frame formulations of de Rham-Gabadadze-Tolley theories}
\label{S2}
\subsection{Metric formulation}
dRGT theories are non-linear Fierz-Pauli theories for which the function $V$ takes a special form.
We will use here the parametrization of dRGT theories proposed in Refs. \cite{Hassan:2011hr,Hassan:2011vm}. We begin by introducing the four functions  $\E_1, \E_2, \E_3, \E_4$ defined for an arbitrary $n \times n$ matrix\footnote{With $a$, a line index belonging to $\{1,...,n\}$, and $b$, a column index belonging to $\{1,...,n\}$, and $n$ having so far no relation to the space-time dimension $D$.} $X^a_{\hphantom{a} b}$, representing elementary symmetric polynomials of the eigenvalues of $X$, and given by (see e.g. \cite{Hassan:2011hr}) 
\ba 
\E_1 \left(X\right) &=& [X] \\
\E_2 \left(X\right) &=& \frac{1}{2} \left([X]^2 - [X^2]\right)\\
\E_3 \left(X\right) &=& \frac{1}{6} \left([X]^3 - 3 [X][X^2] + 2 [X^3]\right) \\
\E_4 \left(X\right) &=& \frac{1}{24} \left([X]^4 - 6[X]^2[X^2] +3[X^2]^2 + 8 [X][X^3]- 6 [X^4]\right)
\ea
where $[X]$ denotes the trace $X^a_{\hphantom{a} a}$ of the matrix $X$. For general $k$, one defines $\E_k$ as 
\ba
\E_k(X) = \frac{1}{k!} X^{a_1}{}_{[a_1}...X^{a_k}{}_{a_k]}, 
\ea
where here and henceforth brackets $[\;]$ denotes the sum over unnormalized antisymmetrized permutations\footnote{and similarly, parentheses $(\;)$ will denote the sum over unnormalized permutations.}. In particular, for a $n \times n$ matrix $X$ one has\footnote{As a simple consequence of Cayley's theorem.} that 
\ba \label{defEn}
{\rm det}  (X) = \E_n(X).
\ea
Furthermore these functions appear in the expansion of the characteristic polynomial of the $D \times D$ matrix $X^\mu_{\hphantom{\mu} \nu}$. Indeed, defining $Y^\mu_{\hphantom{\mu} \nu}$ by 
\ba
Y^\mu_{\hphantom{\mu} \nu} =  X^\mu_{\hphantom{\mu} \nu} - x  \delta^\mu_{\hphantom{\mu} \nu}
\ea
It follows that (here for $D=4$)
\ba  
{\rm det} \left( X^\mu_{\hphantom{\mu} \nu} - x  \delta^\mu_{\hphantom{\mu} \nu}\right) &=& \E_4(Y) \nonumber \\
&=& \sum_{k=0}^{k=4} (-x)^k \E_{4-k} \left(X\right) \nonumber \\
&=& x^4 \E_0(X)- x^3 \E_1(X) + x^2 \E_2(X) -  x \E_3(X) +  \E_4(X) \label{IDENT}
\ea
with the convention that $\E_0= 1$. 
The dRGT theory \cite{deRham:2010kj,deRham:2010ik,deRham:2011rn} can now be defined by an action of the form \cite{Hassan:2011hr} 
\ba \label{dGTBETA}
S = M_P^2 \int d^4 x \sqrt{-g}\left[R -  m^2 \sum_{k=0}^{k=4} \beta_k \E_k\left(\sqrt{g^{-1} f}\right)\right]
\ea
where the $\beta_n$ are arbitrary parameters. 
Note that $\beta_0$ just parametrizes a mere cosmological constant $\Lambda$, which by itself does not give any mass to the graviton, hence in the following we will generally trade $\beta_0$ for $\Lambda$. It might nonetheless be necessary to keep a non vanishing $\beta_0$ in order to have Minkowski space-time as a solution (and fullfill condition (i) of the previous section). On the other hand, the highest order term proportional to $\beta_4$ gives no contribution to the field equations of $g_{\mu \nu}$, 
since $\sqrt{-g}\; \E_4\left(\sqrt{g^{-1} f}\right) = \sqrt{-g} \;{\rm det} \left(\sqrt{g^{-1} f}\right) = \sqrt{-f}$. 
Hence, in $D=4$ dimensions, there is a three  parameter family of non trivial theories, indexed by parameters $\beta_n$, with $n=1,2,3$.
This can easily be extended to $D$ arbitrary dimensions by considering actions of the form 
\ba \label{dGTBETAd}
S = M_P^{D-2} \int d^D x \sqrt{-g}\left[R - 2\Lambda -  m^2 \sum_{k=1}^{k=D-1} \beta_k \E_k\left(\sqrt{g^{-1} f}\right)\right]
\ea
where $\E_n$ are defined as in (\ref{defEn}).
Note that the above definitions (\ref{dGTBETA})-(\ref{dGTBETAd}) use a real matrix square root of the tensor ${\cal M} \equiv g^{-1}f$. In general, however, there is no reason for this square root to exist for arbitrary metrics $g_{\mu \nu}$ and $f_{\mu \nu}$ (see e.g. \cite{sqrt} and \cite{usbis}), and hence one has to assume that it does exist for the above definitions to make sense (we will come back to this issue later). When it does 
we define this square root as $\gamma$, and write 
\ba \label{gammadef} 
\gamma^\mu_{\hphantom{\mu} \sigma}\gamma^\sigma_{\hphantom{\sigma} \nu} = g^{\mu \sigma} f_{\sigma \nu},
\ea
such that one has (for $D=4$ dimensions)
\ba \label{dGTBETAbis}
S = M_P^2 \int d^4 x \sqrt{-g}\left[R - 2\Lambda- m^2 \sum_{k=1}^{k=3} \beta_k \E_k\left(\gamma \right)\right]\ .
\ea

An alternative formulation to action (\ref{dGTBETA}) is to consider $\gamma$ in (\ref{dGTBETAbis}) as an independant field and to enforce the relation (\ref{gammadef}) by a Lagrange multiplier $\cc_{\mu}^{\hphantom{\mu} \nu}$, adding to the Lagrangian a term of  the form (in the spirit of e.g. Ref. \cite{deRham:2010kj})
\ba \label{conslag}
 \sqrt{-g} \cc_{\mu}^{\hphantom{\mu} \nu}\left(\gamma^\mu_{\hphantom{\mu} \sigma}\gamma^\sigma_{\hphantom{\sigma} \nu} - g^{\mu \sigma} f_{\sigma \nu}\right).
 \ea
 This alternative does not feature the presence of the unpleasant square root in the action. 
 Whichever way is chosen (i.e. (\ref{dGTBETA}) or (\ref{dGTBETAbis}) together with (\ref{conslag})) the presence of the square root directly in the action or via a Lagrange multiplier is a somewhat inelegant aspect of the theories considered. As we will show, the vierbein formulation of these theories (or at least of a subset of them) offers a nice alternative which does not suffer from the same lack of elegance.

For future reference, we also define actions $S_\kappa$ by  
\ba \label{SYM}
S_\kappa &=&  M_P^2 \int d^4 x \sqrt{-g}\left[R -  m^2 {\rm det }\left(\kappa \sqrt{g^{-1} f}\;^\mu_{\hphantom{\mu} \nu} -  \delta^\mu_{\hphantom{\mu} \nu}\right)\right] \\
&=& M_P^2 \int d^4 x \sqrt{-g}\left[R - m^2 {\rm det }\left(\kappa \gamma^\mu_{\hphantom{\mu} \nu} -  \delta^\mu_{\hphantom{\mu} \nu}\right)\right] \nonumber \\
&=&  M_P^2 \int d^4 x \sqrt{-g}\left[R -  m^2 \left( \E_0(\gamma)- \kappa \E_1(\gamma) + \kappa^{2} \E_2(\gamma) -  \kappa^{3}  \E_3(\gamma) + \kappa^4 \E_4(\gamma)\right)\right]\nonumber
\ea
where $\kappa$ is a dimensionless parameter (and we have used identity (\ref{IDENT})). 
It is easy to see that on taking linear combinations of such models (with different values of $\kappa$), one can obtain any model (\ref{dGTBETA}) with arbitrary coefficients $\beta_n$ (using the non vanishing of a Vandermonde determinant).

\subsection{Vielbein formulation}
\subsubsection{Generalities}
In order to formulate the dRGT theories in $D$ dimensions using vielbeins, let us define $\thetavierbein^{A}$ and $\fvierbein^{A}$ to be two bases of 1-forms obeying at every space-time point \footnote{Our convention is that Greek letters denote space-time indices, while capital Latin letters denote Lorentz indices that are moved up and down with the canonical Minkowski metric $\eta_{AB}$. We will, however, also use a particular coordinate system in which the $f_{\mu \nu}$ metric just takes the canonical Minkowski $\eta_{\mu \nu}$ form and the vierbeins $\fvierbein^{A}_{\hphantom{A}\mu}$ have components $\delta^{A}_{\hphantom{A}\mu}$. In that case it sometimes turns out to be convenient to use the same type of letter to denote Lorentz and space-time indices - one then has to pay attention at the order of the indices to be able to discriminate between them.}
\be
g^{\mu\nu}\thetavierbein^{A}_{\hphantom{A}\mu}\thetavierbein^{B}_{\hphantom{B}\nu}=f^{\mu\nu}\fvierbein^{A}_{\hphantom{A}\mu} \fvierbein^{B}_{\hphantom{B}\nu}=\eta^{AB}\ ,
\ee
or equivalently
\ba
&\eta_{AB}\thetavierbein^{A}{}_{\mu}\thetavierbein^{B}{}_{\nu}=g_{\mu\nu}\ ,\\
&\eta_{AB}\fvierbein^{A}{}_{\mu}\fvierbein^{B}{}_{\nu}=f_{\mu\nu}\ .
\ea
We will also need the vectors $e_{A}$ and $\fmoinsunvierbein_{A}$, respectively dual to the 1-forms $\thetavierbein^{A}$ and $\fvierbein^A$, that verify
\ba \thetavierbein^{A}(e_{B})=\thetavierbein^{A}{}_{\mu}e_{B}{}^{\mu}=\delta^{A}{}_{B}, \\
\fvierbein^{A}(\ell_{B})=\fvierbein^{A}{}_{\mu}\fmoinsunvierbein_{B}{}^{\mu}=\delta^{A}{}_{B}.
\ea
For future use, we define the $(D-n)-$forms $\thetavierbein^*_{A_1\dots A_n}$  (using the notations of Ref.~\cite{DuboisViolette:1986ws}) by   
\footnote{Notice that the $(D-n)-$forms $\thetavierbein^*_{A_1\dots A_n}$ carry $n$ Lorentz indices and that this definition also makes sense for $n=0$ in which case the form $\thetavierbein^*$ is just proportional to the volume $D$-form.}
\be
\thetavierbein^*_{A_1\dots A_n}\equiv{1\over (D-n)!}\epsilon_{A_1\dots A_D}
\thetavierbein^{A_{n+1}}\wedge\dots\wedge \thetavierbein^{A_{D}}\ ,
\ee
where $\epsilon_{A_{1}\dots A_{D}}$ is the totally antisymmetric tensor verifying $\epsilon_{123\dots D}=1$. It then follows that
\be
\thetavierbein^{A}\wedge\thetavierbein^{*}_{A_{1}\dots A_{n}}=\sum_{k=1}^{n}(-1)^{n-k}\delta^{A}{}_{A_{k}}\thetavierbein^{*}_{A_{1}\dots A_{k-1}A_{k+1}\dots A_{n}}\ .
\ee

Using the vielbeins $\thetavierbein^{A}$ the Einstein-Hilbert action for the dynamical metric $g_{\mu \nu}$ reads 
\be
S_{EH}=M_P^{D-2}\int \Omega^{AB}\wedge\thetavierbein^*_{AB},
\label{action}
\ee
where 
\be
\Omega^{AB}\equiv d\omega^{AB}+\omega^{A}{}_C\wedge\omega^{CB}\ ,
\ee
is the curvature 2-form associated with the spin-connection $\omega^{AB}$. The latter is a one form taking values in the Lie Algebra of $SO(1,3)$ (hence it is antisymmetric in its Lorentz indices $A, B$) and can be expressed in terms of the vielbein $\thetavierbein^{A}$ assuming (as we shall do henceforth) the torsion free condition \footnote{We recall that this condition can also be obtained by considering the spin connection as an independent field, and writing its field equations.} 
\ba 
\mathcal{D}\thetavierbein^A= d\thetavierbein^A+\omega^{A}{}_B\wedge\thetavierbein^B=0,
\label{TORSION}
\ea
where the derivative operator $\mathcal{D}$ acting on an arbitrary $p$-form carrying Lorentz indices $\Pi^{A_1...A_n}_{\hphantom{A_1...A_n}B_1...B_m}$ is defined by 
\ba
{\mathcal D} \Pi^{A_1\cdots A_n}_{\hphantom{A_1\cdots A_n}B_1 \cdots B_m} = d \Pi^{A_1 \cdots A_n}_{\hphantom{A_1 \cdots A_n}B_1 \cdots B_m} \nonumber 
 &+& \sum_{p=1}^{p=n} \omega^{A_p}{}_C \wedge \Pi^{A_1 \cdots A_{p-1} C A_{p+1} \cdots A_n}_{\hphantom{A_1 \cdots A_{p-1} C A_{p+1} \cdots A_n}B_1 \cdots B_m} \nonumber \\
&-& \sum_{p=1}^{p=m} \omega^{C}{}_{B_p}\wedge \Pi^{A_1 \cdots  A_n}_{\hphantom{A_1 \cdots  A_n}B_1 \cdots B_{p-1} C B_{p+1} \cdots B_m}.
\ea
Using (\ref{TORSION}), one finds easily the components of the spin connection $\w_{ABC}$ which are defined by one of  the two equivalent relations 
\ba 
\w_{ABC} &=&   e_C^{\hphantom{A}\mu} \omega_{A B \mu}, \\
\omega_{A B \mu} &=& E^C{}_\mu \w_{ABC}, \label{defw}
\ea
and are given by  
\be
\begin{split}
\w_{ABC}=&\frac{1}{2}(e_{B}{}^{\mu}e_{C}{}^{\nu}\partial_{\mu}\thetavierbein_{A\nu}-e_{C}{}^{\mu}e_{B}{}^{\nu}\partial_{\mu}\thetavierbein_{A\nu}+e_{C}{}^{\mu}e_{A}{}^{\nu}\partial_{\mu}\thetavierbein_{B\nu}\\
&-e_{A}{}^{\mu}e_{C}{}^{\nu}\partial_{\mu}\thetavierbein_{B\nu}-e_{A}{}^{\mu}e_{B}{}^{\nu}\partial_{\mu}\thetavierbein_{C\nu}+e_{B}{}^{\mu}e_{A}{}^{\nu}\partial_{\mu}\thetavierbein_{C\nu}),
\label{connection}
\end{split}
\ee
while the components of the curvature read 
\ba
\W^{AB}{}_{CD} &\equiv & \Omega^{AB}{}_{\mu \nu} e_C{}^{\mu} e_D{}^{\nu} \nonumber \\
 &=&
e_{[C}{}^\mu \partial_\mu(\w^{AB}{}_{D]})+\w^{A}{}_{E[C}\w^{EB}{}_{D]} -\w^{AB}_{\hphantom{AB}G}\thetavierbein^G{}_{\nu} e_{[C}{}^\mu \partial_\mu e_{D]}{}^\nu\ .
\label{curvature}
\ea 
The curvature satisfies the Bianchi identities 
\be
\mathcal{D}\Omega^{AB}\equiv d\Omega^{AB}+\omega^{A}{}_{C}\wedge\Omega^{CB}+\omega^{B}{}_{C}\wedge\Omega^{AC}=0\ .
\ee
The Einstein tensor has a simple expression in terms of the $\thetavierbein^*_{ABC}$ forms and is given by the  $D-1$ form 
\be
G_A\equiv \frac{1}{2}\Omega^{BC}\wedge\thetavierbein^*_{ABC}.
\ee
Furthermore it obeys Bianchi identities
\be \label{BIANCHI}
\mathcal{D}G_A=0\ ,
\ee
which involve derivative of the $D-1$ form $G_A$, and hence just yield $D$ coordinate-scalar equations. Additionally, because of the local Lorentz invariance of the Einstein-Hilbert term, the Einstein tensor decomposed as $G_A\equiv G_{A}{}^{B}\thetavierbein^*_B$ satisfies 
\be \label{SYMGAB}
G_{[AB]}=0,
\ee
i.e. $G_{AB}$ is symmetric. 

\subsubsection{Mass terms and field equations}
As discussed above the mass terms of the dRGT theory are expressed in terms of the (matrix) square root $\gamma$ of $g^{-1}f$.
Defining the Lorentz tensors $\hat{e}_{A}{}^{B}$ (whose indices are moved up and down with $\eta_{AB}$) as 
\ba 
\hat{e}_{A}{}^{B}=e_{A}{}^{\mu}\fvierbein^{B}{}_{\mu},
\ea
 a sufficient condition for this square root to exist is that the vierbeins obey the condition \cite{Chamseddine:2011mu,Volkov:2012wp} (see also \cite{usbis})
\ba \label{CONS} 
\hat{e}^{AB} = \hat{e}^{BA},
\label{hatsymeab}
\ea
in which case, $\gamma$ defined as  
\ba \label{gammavier}
\gamma^\mu_{\hphantom{\mu} \nu} = e^{\hphantom{A}\mu}_A \fvierbein^A_{\hphantom{A}\nu}
\ea
verifies (\ref{gammadef}). 
Note that whenever the non dynamical metric $f_{\mu \nu}$ is flat, a convenient choice of vierbein $\fvierbein^A$ can be made by first choosing cooordinates $x^\mu$ where  $f_{\mu \nu}$ takes the canonical form $\eta_{\mu \nu}$, i.e. 
\ba \label{GAUG1}
f_{\mu \nu} = \eta_{\mu \nu},
\ea
and then choosing $\fvierbein^A = dx^A$, i.e. such that (in components)
\ba \label{GAUG2}
\fvierbein^{A}_{\hphantom{A}\mu} = \delta^{A}_{\hphantom{A}\mu}.
\ea 
In that case it is sometimes convenient to label space-time indices and Lorentz indices with the same set of letters, which we will do using latin capital letters. When the choice (\ref{GAUG1})-(\ref{GAUG2}) is made, 
one has 
\ba
\hat{e}_{A}{}^B = e_{A}{}^B
\ea
and the constraint (\ref{CONS}) simply reads 
\ba \label{symeab}
e^{AB} = e^{BA},
\ea
stating that the vierbein $e^{A \mu}$ can be represented as a symmetric matrix.

On substituting the expression (\ref{gammavier}) into the mass terms of Eq.(\ref{dGTBETA}) it follows that these can be rewritten in terms of the vierbeins as 
\ba
M_P^2 m^2 \sum_{n=0}^{4} \beta_n\int \fvierbein^{A_1}\wedge\dots\wedge \fvierbein^{A_n}\wedge\thetavierbein^*_{A_1\dots A_n}\ ,
\label{massvier}
\ea
where we have absorbed irrelevant numerical coefficients $n!$ by redefining the $\beta_n$.
Also note that using the same substitution (and $\sqrt{-g} = {\rm det} (\thetavierbein^A_{\hphantom{A}\mu}) \equiv \thetavierbein$), the action $S_{\kappa}$ of Eq. (\ref{SYM}) now reads
\ba \label{detx}
S_\kappa &=& M_P^2 \int d^4 x \thetavierbein \left[R -  m^2 {\rm det }\left(\kappa e^{\hphantom{A} \mu}_A \fvierbein^A_{\hphantom{A}\nu} -  \delta^\mu_{\hphantom{\mu} \nu}\right)\right] \\
&=& M_P^2 \int d^4 x \left[ \thetavierbein  R -  m^2 {\rm det} \left( \kappa \fvierbein^A_{\hphantom{A}\nu} - \thetavierbein^A_{\hphantom{A}\nu}\right)\right]. \label{detxbis}
\ea 
As a side remark, we note that from action (\ref{detxbis}) one can extract the kinetic terms for the extra polarization of a massive graviton in a very simple way i.e. by a simple shift of the vierbein. As far as we know, this trick has not been noticed before and differs from Stuckelberg's. However, since this is not the main subject of this paper, we discuss it in more detail in appendix \ref{appC}.

Hence, generalizing to $D$ dimensions, one is led to consider theories defined by the action 
\be
S=M_P^{D-2} \int \Omega^{AB}\wedge\thetavierbein^*_{AB}-M_P^{D-2} m^2 \sum_{n=0}^{D-1}\beta_n\int \fvierbein^{A_1}\wedge\dots\wedge \fvierbein^{A_n}\wedge\thetavierbein^*_{A_1\dots A_n}\ ,
\label{action}
\ee
where the kinetic Einstein-Hilbert term can easily be generalized to the Gauss-Bonnet-Lovelock terms in  $D$ dimensions.
 
  Varying this action with respect to the forms $\thetavierbein^A$ we get the following field equations in vacuum (see appendix \ref{appA} for a derivation) 
\be
G_A= t_A\ ,
\label{eom}
\ee
with $t_A$ defined by 
\be \label{defta}
t_A\equiv \frac{1}{2}\sum_{n=0}^{D-1} \beta_n \fvierbein^{A_1}\wedge\dots\wedge \fvierbein^{A_n}\wedge\thetavierbein^*_{AA_1\dots A_n}\equiv
t_{A}{}^{B}\thetavierbein^*_B\ ,
\ee
and we have set $m^2$ to one for convenience.
Using 
\ba \label{defAB}
\fvierbein^{A}\equiv \hat{e}_{B}{}^{A}\thetavierbein^{B}, 
\ea
the coefficients $t_{A}{}^{B}$ can be computed to be  
\be
t_{A}{}^{B}=\frac{1}{2}\sum_{n=0}^{D-1}\beta_{n}\hat{e}_{B_{1}}{}^{A_{1}}\dots \hat{e}_{B_{n}}{}^{A_{n}}\delta^{BB_{1}\dots B_{n}}_{AA_{1\dots A_{n}}}\ ,
\label{emt}
\ee
where
\ba
\delta_{A A_1\dots A_n}^{B B_1\dots B_n}&\equiv&\delta_{A}^{[B} \delta_{A_1}^{B_{1}}\dots \delta_{A_{n}}^{B_n]}, \\
&=& \frac{1}{(D-n-1)!}\epsilon^{B B_1 \dots B_{n}C_{1}\dots C_{D-n-1}}\epsilon_{A A_{1}\dots A_{n}C_{1}\dots C_{D-n-1}}.
\label{formula}
\ea

Note that the above set of theories (\ref{action}) (with field equations (\ref{eom})) are perfectly well defined 
without imposing the constraint (\ref{CONS}). Furthermore we will show  that in some cases,
the constraint (\ref{CONS}) arises as a consequence of the field equations \footnote{Notice that this  dynamically enforces the existence of the square root $\gamma$, a nice feature of the vielbein formulation.}. That this is the case has already been argued in Ref. \cite{Hinterbichler:2012cn}. However, the arguments presented in Ref. \cite{Hinterbichler:2012cn} use a decomposition of a generic vierbein 
 (also adopted in Ref. \cite{Nibbelink:2006sz}) as a product of a Lorentz transform with a symmetric vierbein. More specifically they assume the validity of the ``Minkowski" version of the polar decomposition, assuming an arbitrary invertible matrix $\mathrm{m}$ can always be written as the product of a Lorentz transform $\mathrm{\lambda}$ with a symmetric matrix $\mathrm{s}$, 
 \ba \label{Polarbis}
\mathrm{m}= \mathrm{\lambda} \mathrm{s}.
\ea
However, it can be shown (see e.g. \cite{usbis}) that this decomposition only holds for a restricted set of matrices $\mathrm{m}$, so that the arguments presented in \cite{Hinterbichler:2012cn} are in fact not fully conclusive. Notice that one has argued that the constraint (\ref{CONS}) 
can be set by a suitable Lorentz gauge choice \cite{Volkov:2012wp}. In fact this argument also uses a decomposition such as (\ref{Polarbis}), and so  is not always valid. This will be discussed in detail elsewhere \cite{usbis}. 

Hence, in the following we use action  (\ref{action}) as a starting point, and  
 begin by dropping the constraint (\ref{CONS}). As such, the set of theories (\ref{action}) just defines some vielbein theory with a non dynamical vielbein $\fvierbein^A$ and an unconstrainded dynamical vielbein $\thetavierbein^A$, carrying then, for $D=4$, 16 a priori dynamical polarizations.

\section{Counting degrees of freedom using vielbeins}
\label{S4}
In this section, we will successively see how various constraints reduce the number of physical polarizations of the {\it a priori} unconstrained vielbein. We stress that most of our results will be valid for any $D$ space-time dimensions. For $D=4$, the constraints discussed in the following two subsections allow respectively to go from 16 to 10 dynamical components and then from 10 to 6. We will show in the third subsection below that an extra scalar constraint can be obtained for a subset of theories.

\subsection{Constraints arising from local Lorentz invariance}
The constraints arising from local Lorentz
invariance are encoded in the symmetry of $G_{AB}$ (see Eq. (\ref{SYMGAB})) and read, from the field equations (\ref{eom}) 
\be \label{SYMETRICTAB}
t_{[AB]}=0\ .
\ee
There are $D(D-1)/2$ independent constraints arising from (\ref{SYMETRICTAB}). One therefore expects those to restrict the number of independent components of the dynamical vielbein $e_{B}^{\hphantom{B}\mu}$, a priori $D \times D$ (i.e. 16 for $D=4$ dimensions), to $D^2 - D(D-1)/2 = D(D+1)/2$ which is the same number of components as in the metric $g_{\mu \nu}$ (i.e. $10$ for $D=4$ dimensions). We will even show that in some cases, constraints (\ref{SYMETRICTAB}) further simplify, turning out to be equivalent to the conditions (\ref{hatsymeab}).

First, it is easy to see that assuming $\hat{e}^{AB}$ symmetric is sufficient to yield a symmetric $t^{AB}$ (or equivalently $t_{AB}$). However,  the converse is not true in general. For example the term proportional to $\beta_2$ in the right hand side of Eq. (\ref{emt}) yields a symmetric contribution to $t^{AB}$ when $\hat{e}^{AB}$ is assumed {\it antisymmetric}. Of course,   Eq. (\ref{hatsymeab}) can always be imposed (if needed) by introducing it by hand in the theory, i.e. by suitable Lagrange multipliers. As we just stressed, this will be compatible with field equations.

There are, however, cases in which the equivalence can be established. 
In particular when only $\beta_{D-1}\neq 0$ (besides possibly $\beta_0$, since the contribution of the latter in  (\ref{emt}) yields always a symmetric $t^{AB}$) one has 
\be
t_{A}{}^{B}\propto \hat{e}_{B_{1}}{}^{A_{1}}\dots \hat{e}_{B_{d-1}}{}^{A_{d-1}}\delta^{BB_{1}\dots B_{D-1}}_{AA_{1}\dots A_{D-1}}=\det(\fvierbein)\hat{\thetavierbein}^{B}{}_{A}\ ,
\ee
where the matrix $\hat{\thetavierbein}$ is just the transposed inverse of $\hat{e}$ and is defined by its coefficients $\hat{\thetavierbein}^{B}{}_{A}$ satisfying 
\be
\fmoinsunvierbein_{A}\equiv \hat{\thetavierbein}^{B}{}_{A}e_{B}.
\ee
Notice that with the choices (\ref{GAUG1})-(\ref{GAUG2}) one has 
\be
\hat{\thetavierbein}^{B}{}_{A}=\fmoinsunvierbein_{A}{}^{\mu}\thetavierbein^{B}{}_{\mu}=\delta^{\mu}{}_{A}\thetavierbein^{B}{}_{\mu}=\thetavierbein^{B}{}_{A}\ .
\ee
In this case (non vanishing $\beta_{D-1}$), the symmetry of $t_{AB}$ implies the symmetry of $\hat{\thetavierbein}_{AB}$ (because $\det(\fvierbein)\neq 0$). This automatically implies that $\hat{e}^{AB}$ is also symmetric.
This also happens when only $\beta_{1}\neq 0$, in this case one has 
\be
t_{A}{}^{B}\propto \hat{e}_{B_{1}}{}^{A_{1}}\delta_{AA_{1}}^{BB_{1}}=\text{tr}(\hat{e})\delta_{A}^{B}-\hat{e}_{A}{}^{B}\ ,
\ee
which again yields a symmetric $\hat{e}^{AB}$. For an arbitrary combination of mass terms with non vanishing $\beta_n$ no definite conclusion can be drawn, however choosing suitable $\beta_{n}$ results in the mass term appearing in Eq. (\ref{detxbis})  and yields (with the choices (\ref{GAUG1})-(\ref{GAUG2}))
\be
t_{A}{}^{B}\propto \frac{\partial\det(\kappa \hat{e}-\text{Id})}{\partial \hat{e}_{B}{}^{A}}=\det(\kappa \hat{e}- \text{Id})[(\kappa \hat{e}- \text{Id})^{-1}]^{B}{}_{A}\ ,
\ee
which includes the two previous cases as limiting cases (when $\kappa=0$ and $\kappa\rightarrow\infty$).

Notice that whenever we choose the gauge (\ref{GAUG2}), we have $\hat{e}^{AB}=e^{AB}$ and $\hat{E}_{AB}=E_{AB}$,  and thus the symmetry of the hatted quantities implies the symmetry of the vielbeins themselves.
\label{symfab}

\subsection{Constraints arising from diffeomorphism invariance}
\label{10to6}
Using the Bianchi identity (\ref{BIANCHI}) in the field equations we get that  
\be
\mathcal{D}t_A=0,
\label{constraint}
\ee
 which reads explicitly 
\be
\mathcal{D}t_{A}=\frac{1}{2} \sum_{n=1}^{D-1}n \beta_n \mathcal{D}\fvierbein^{A_1}\wedge\dots\wedge \fvierbein^{A_n}\wedge\thetavierbein^*_{AA_1\dots A_n}=0.
\ee
Here, and for the remaining of this subsection and the next one, we choose the gauge (\ref{GAUG2}) which results in the vanishing of $d\fvierbein^{A}$, the above equation becomes
\be
\sum_{n=1}^{D-1}n \beta_n\ \omega^{A_{1}}{}_{B}\wedge \fvierbein^{B}\wedge \fvierbein^{A_2}\wedge\dots\wedge \fvierbein^{A_n}\wedge\thetavierbein^*_{AA_1\dots A_n}=0. 
\label{constraintgen}
\ee
On using $\fvierbein^{A_{i}}\equiv \hat{e}_{B_{i}}{}^{A_{i}}\ \thetavierbein^{B_{i}}=e_{B_{i}}{}^{A_{i}}\ \thetavierbein^{B_{i}}$, $\omega^{A_1}{}_{B}\equiv \w^{A_1}{}_{B\,\, C}\ \thetavierbein^C$  and the identity (deduced from (\ref{formula}))
\be
\thetavierbein^{B_{1}}\wedge\dots\wedge\thetavierbein^{B_{n}}\wedge\thetavierbein^{*}_{A_{1}\dots A_{n}}=\varepsilon \delta^{B_{1}\dots B_{n}}_{A_{1}\dots A_{n}},
\label{formula2}
\ee
where $\varepsilon$ denotes the volume $D$-form $\thetavierbein^1\wedge \thetavierbein^2 \wedge \cdots \wedge \thetavierbein^D$,
we can rewrite (\ref{constraintgen})  in the useful form 
\be
\sum_{n=1}^{D-1}n\beta_{n}\ \w^{A_1}{}_{B\,\, C}e_{B_1}{}^{B}e_{B_2}{}^{A_2}\dots e_{B_n}{}^{A_n}
\delta_{A A_1\dots A_n}^{C B_1\dots B_n} \varepsilon=0 .
\label{constraintbis}
\ee
 Constraints (\ref{constraint}) clearly remove another $D$ degrees of freedom. So, in $D=4$ dimensions we have so far shown that from the $16$ components only $16-6-4=6$ are left dynamical and one needs one more constraint to have only the sought for $5$ dynamical degrees of freedom of a massive graviton. In the following, specializing to $D=4$ dimensions, we will discuss how such a constraint can arise in some specific cases.

\subsection{Extra scalar constraint}
\label{6to5}
In this section we particularize our discussion to $D=4$ dimensions and to the gauge choice (\ref{GAUG2}).
The way we proceed to obtain the extra constraint is very similar to what was done in the quadratic Fiez-Pauli theory, as explained in section \ref{S1}, where the extra constraint is given by Eq. (\ref{CONS2}). Namely one uses the constraints coming from the Bianchi identity (\ref{constraint}) (cf. Eq. (\ref{EQAUX1})) back into a suitable trace of the field equations  
(\ref{eom}) (cf. Eq. (\ref{FIELDgen})), which in our case reads
\be
m^{A}\wedge G_A=m^A\wedge t_A,
\label{claim}
\ee
where $m^A$ is a suitable collection of one forms (labelled by the Lorentz index $A$). As we will now show, however, things proceed differently depending on the values of the $\beta_n$ coefficients.

\subsubsection{Only $\beta_{1}\neq 0$} \label{b1}
We first discuss the case in which the only non vanishing coefficient $\beta_n$ is $\beta_1$ (and possibly $\beta_0$ parametrizing then a non vanishing cosmological constant), because this case is the simplest. 
Then, we can rewrite the Bianchi identity (\ref{constraintbis})  as (the term proportional to $\beta_0$ automatically vanishes)
\be
\begin{split}
0=\mathcal{D}t_{A}&=\frac{1}{2}\beta_{1}\ \w^{A_{1}}{}_{B\ C}e_{B_{1}}{}^{B}\delta^{CB_{1}}_{AA_{1}}\varepsilon\\
&=\frac{1}{2}\beta_{1}\ \w^{B}{}_{CB}e_{A}{}^{C}\varepsilon,
\label{n=1}
\end{split}
\ee
which immediately implies
\ba 
\w^{B}{}_{AB}=0,
\ea  yielding then 
\ba \label{consn1}
e_{B}{}^{C}\partial_{C}E^{B}{}_{A}-e_{B}{}^{C}\partial_{A}E^{B}{}_{C}=0.
\ea
Choosing  $m^A =\thetavierbein^A$ and tracing over the field equations (\ref{eom}) we get 
\be \label{finaln1}
\thetavierbein^{A}\wedge G_{A}=\thetavierbein^{A}\wedge t_{A}.
\ee
The left hand side of the above equation can be rewritten using  (\ref{curvature}) and (\ref{formula2}) as
\ba
\thetavierbein^{A}\wedge G_{A}&=&\frac{1}{2}\thetavierbein^{A}\wedge\Omega^{BC}\wedge\thetavierbein^{*}_{ABC},\nonumber \\
&\sim& \varepsilon \W^{BC}{}_{DE}\ \delta^{DE}_{AB}  , \nonumber \\&\sim& \varepsilon e_{A}{}^\mu \partial_\mu(\w^{BA}{}_{B})\ ,
\ea
where the symbol $\sim$ means that we only write terms containing second order derivatives and omit an overall constant factor. But the constraint (\ref{consn1}) tells us that these terms vanish, so equation (\ref{finaln1}) contains only derivatives of order one at most, and hence represents an additional constraint. 
This constraint, together with the four vector constraints (\ref{consn1}) can be elegantly rewritten using the following decomposition of the vierbein (valid only on the ``branch" of vierbeins with positive determinants)
\be
E^{A}\equiv e^{\sigma}\bar{E}^{A}
\label{CONF}
\ee 
with $\det(\bar{E})=1$, which immediately implies
\ba
E^{A}{}_{B}= e^{\sigma}\bar{E}^{A}{}_{B}\ , \\
e_{B}{}^{A}= e^{-\sigma}\bar{e}_{B}{}^{A}.
\ea
Now the vector constraint (\ref{consn1})  becomes 
\be
\partial_{A}\sigma=\frac{1}{3}\bar{e}_{B}{}^{C}\partial_{[C}\bar{E}^{B}{}_{A]}\,
\ee
which, using the fact that the matrix whose matrix elements are the $\bar{E}^{A}{}_{B}$ has a unit determinant (which implies that $\bar{e}_{B}{}^{C}\partial_{A}\bar{E}^{B}{}_{C}$ vanishes) can be rewritten as 
\be
\partial_{A}\sigma=\frac{1}{3}\bar{e}_{B}{}^{C}\partial_{C}\bar{E}^{B}{}_{A}\ .
\ee
Eq. (\ref{CONF}) implies on the other hand that 
\be
e^{\sigma}\w_{ABC}=\wb_{ABC}-\bar{e}_{[A}{}^{\mu}\partial_{\mu}\sigma\ \eta_{B]C}\ ,
\label{CONFCON}
\ee
where  the coefficients $\wb_{ABC}$ are defined as those, $\w_{ABC}$, of $\omega_{AB}$ (see Eq.(\ref{defw})) by  $\bar{\omega}^{A}{}_{B}\equiv\wb^{A}{}_{BC}\bar{E}^{C}$. In terms of those coefficients, the vector constraints (\ref{consn1}) just read 
\be
3\bar{e}_{A}{}^{B}\partial_{B}\sigma=-\partial_{B}\bar{e}_{A}{}^{B}=-\wb^{B}{}_{AB}.
\label{BIANCHI1}
\ee
On the other hand, using (\ref{CONF}) and (\ref{CONFCON}) the scalar constraint (\ref{finaln1}) just reads
\be
-\wb_{ABC}\wb^{BCA}+\frac{1}{3}\wb^{B}{}_{AB}\wb^{CA}{}_{C}+2\wb^{AB}{}_{C}\bar{e}_{A}{}^{H}\bar{e}_{B}{}^{G}\partial_{H}\bar{\thetavierbein}^{C}{}_{G}
=2\beta_{0}e^{2\sigma}+\frac{3}{2}\beta_{1}e^{\sigma}\text{tr}(\bar{e}_{A}{}^{B})\,
\label{GHOSTBUSTER}
\ee
which shows clearly that beyond the first derivative already constrained by the vector constraints another independent Lorentz invariant relation  between first derivatives can be obtained.

\label{b1}

\subsubsection{Only $\beta_{2}\neq 0$}

\label{b2}

The starting point here is the same as above, starting from the vaninshing of ${\mathcal D}t_A$, and using first only the symmetry of $t^{AB}$ which implies 
\ba \label{consomt}
\w^{A}{}_{BC} t_{A}{}^B = 0,  
\ea
we can rewrite  Eq. (\ref{constraintbis}), as 
\ba
0=\mathcal{D}t_{A}&=&\beta_{2}\ \w^{A_{1}}{}_{B\ C}e_{B_{1}}{}^{B}e_{B_{2}}{}^{A_{2}}\delta^{CB_{1}B_{2}}_{AA_{1}A_{2}}\varepsilon\\
&=&\beta_{2}\left(\w^{B}{}_{CB}e_{D}{}^{C}e_{A}{}^{D}+\w^{B}{}_{CD}e_{A}{}^{C}e_{B}{}^{D}\right.\nonumber \\
&& \left. -\w^{B}{}_{CB}e_{A}{}^{C}e_{D}{}^{D}-\w^{B}{}_{CD}e_{B}{}^{C}e_{A}{}^{D}\right)\varepsilon\, 
\label{n=2}
\ea
yielding 
\be
\w^{B}{}_{CB}e_{D}{}^{C}e_{A}{}^{D}+\w^{B}{}_{CD}e_{A}{}^{C}e_{B}{}^{D}-\w^{B}{}_{CB}e_{A}{}^{C}e_{D}{}^{D}-\w^{B}{}_{CD}e_{B}{}^{C}e_{A}{}^{D}=0\ ,
\ee
which reduces to the simpler form 
\ba
\w^{B}{}_{CB}e_{A}{}^{C}+\w^{B}{}_{AC}e_{B}{}^{C}-\w^{B}{}_{AB}e_{C}{}^{C}-\w^{B}{}_{CA}e_{B}{}^{C}=0.\ea 
Due to the form of this equation and to the fact that the left hand side of (\ref{eom}) does not contain any $\fvierbein^A$ it appears natural to use here $m^A= \fvierbein^A$ to trace over the field equations. Using the same logic (and notation) as before we get that  
\ba
\fvierbein^{A}\wedge G_{A}&\sim& \varepsilon e_{A}{}^\mu \partial_\mu (\w^{B}{}_{CB}e^{CA}+\w^{BA}{}_{C}e_{B}{}^{C}-\w^{BA}{}_{B}e_{C}{}^{C})\nonumber\\
&\sim& \varepsilon e_{A}{}^\mu \partial_\mu(\w^{B}{}_{C}{}^{A}e_{B}{}^{C}).
\ea
The right hand side above does not vanish in general, and  hence, in this case the equation $\fvierbein^{A}\wedge G_{A}=\fvierbein^{A}\wedge t_{A}$ does not automatically provide the extra constraint we need. However, if one imposes $e^{AB}$ to be symmetric (which, as we stressed in \ref{symfab}, does not contradict field equations), then $\w^{B}{}_{C}{}^{A}e_{B}{}^{C}$ vanishes (by the same logic as in Eq. (\ref{consomt})), and hence we obtain an extra constraint similar to the one found above. However it should be stressed that this constraint has been obtained using a procedure which differs in its details from the one of \ref{b1}: first we had to trace with a different set of one forms $m^A$ and second, we had to impose by hand the symmetry of $e^{AB}$. 

\subsubsection{Other cases}
A scalar constraint for an arbitrary combination of terms with non vanishing $\beta_0$, $\beta_1$ and $\beta_2$ can easily be obtained along the previous lines. One just needs to trace over the field equations with an appropiate combination of $\fvierbein^{A}$ and $\thetavierbein^{A}$ in order to make all second derivatives disappear. 

However things proceed quite differently whenever $\beta_3$ is non vanishing.  Indeed, let us now discuss the case in which only $\beta_3$ differs from zero. First recall that now there is no need to {\it assume} that $e^{AB}$ is symmetric since it appears as a mere consequence of the field equations. The vector constraints also take a simple form, indeed  we can rewrite (\ref{constraintgen}) as
\ba
0=\mathcal{D}t_{A} &=&\frac{3}{2}\beta_{3}\omega^{A_{1}}{}_{B\mu}dx^{\mu}\wedge dx^{B}\wedge dx^{A_{2}}\wedge dx^{A_{3}}\epsilon_{AA_{1}A_{2}A_{3}} \nonumber\\
&\propto & \beta_{3}\w^{B}{}_{AC}\thetavierbein^{C}{}_{B}\varepsilon. 
\label{n=3}
\ea
Hence, the vanishing of ${\mathcal D}t_{A}$ yields that of  $\w^{B}{}_{AC}\thetavierbein^{C}{}_{B}$ which we can rewrite in terms of the vierbeins as
\be
\partial_{A}\left(\thetavierbein^{B}{}_{B}\right)-\partial_{B}\thetavierbein^{B}{}_{A}=0.
\ee
However in this case, it is not possible (as shown in appendix \ref{appB}) to find a collection of one forms $m^{A}$ such that the equation $m^{A}\wedge G_{A}=m^{A}\wedge t_{A}$ provides an additional constraint in the same way as in the previous cases (at least under some fairly general hypotheses on $m^A$). Hence this case does not appear as transparent as the others as far as the existence of the extra constraint is concerned. The same would be true for an arbitrary linear combination of
mass terms where $\beta_3$ is non vanishing.

\label{b3}

\subsection{Recovering quadratic Fierz-Pauli theory}
It is instructive to show how the constraints of the quadratic Fierz-Pauli theory, Eqs. (\ref{CONS1})-(\ref{CONS2}) can be recovered from the constraints derived in sections \ref{10to6} and \ref{6to5} by expanding the non linear massive gravity around a flat space-time. This will in turn allow us to show that the constraints we obtained are independant from each-other, as they should be. 
To do so, we first need to make sure that flat space-time is indeed a solution of massive gravity and provides a suitable background which might require  adding a non vanishing cosmological constant. This addition does however not change any of the conclusion of the previous sections, as we already stressed. 

Let us then look at the linearized limit of the constraints. We expand the dynamical vierbein as 
\ba
\thetavierbein^{A}&=&dx^{A}+\thetavierbein_{(1)}^{A},\\
e_{A}&=&\partial_{A}+e^{(1)}_{A}, 
\ea
where here and henceforth an index ${}_{(1)}$ denotes a first order perturbation. Writing the metric perturbation as $h_{\mu \nu}$, which verifies $g_{\mu\nu}=\eta_{\mu\nu}+h_{\mu\nu}$, we have that 
\ba
&\thetavierbein^{(1)}_{(AB)}&=h_{AB},\\
&e_{(1)}^{(AB)}&=-h^{AB},\\
&\thetavierbein_{(1)}^{A}{}_{B}&=-e^{(1)}_{B}{}^{A},\
\ea
where, because we are working at linear order, Lorentz and space-time indices are identified. Choosing then $\fvierbein^A$ as in (\ref{GAUG2}), and  writing 
\ba
\hat{e}_{B}{}^{A}=\delta^{A}{}_{B}+\hat{e}^{(1)}_{B}{}^{A},
\ea
we get that 
\be
\hat{e}_{(1)}^{(AB)}=e_{(1)}^{(AB)}=-h^{AB}\ .
\ee
Now at linear order $t_{A}{}^{B}$ reads  ${t}^{(1)}_{A}{}^{B}\propto \text{Tr}(\hat{e}_{(1)})\delta^{B}{}_{A}-\hat{e}^{(1)}_{A}{}^{B}$ , which implies that $\hat{e}_{(1)}^{BA}$ is symmetric and therefore that 
\be
\thetavierbein^{(1)}_{AB}=\frac{h_{AB}}{2}\ .
\ee
In order to be able to write the constraints in terms of $h_{AB}$ we also need the expression of the connection 1-form. From (\ref{connection}) we get
\be
\w^{(1)}_{ABC}=\frac{1}{2}(\partial_{B}h_{AC}-\partial_{A}h_{BC})\ .
\ee
It is now easy to see that, at linear order, the constraints (\ref{constraintbis}) are just
\be
\w_{(1)}^{B}{}_{AB}=\frac{1}{2}(\partial_{A}h^{B}{}_{B}-\partial^{B}h_{AB})=0\ ,
\ee
which is nothing but the vector Pauli-Fierz constraint (\ref{EQAUX1}). 

Let us now examine the  additional scalar constraint corresponding to the scalar Fierz-Pauli constraint (\ref{CONS2}). 
To do this we  look at the trace (\ref{claim}), where $m^A$ is either  $\thetavierbein^A$ or $\fvierbein^A$ (or a combination thereof). 
 As we would expect, at the linear level, the left hand side of Eq.(\ref{claim}) vanishes 
\be
m^{A}\wedge G_{A}\propto \partial_{A}(\w_{(1)}^{BA}{}_{B})\varepsilon = \frac{1}{2}\partial^{A}(\partial_{A}h^{B}{}_{B}-\partial^{B}h_{AB})\varepsilon=0 ;
\ee 
as for the right hand side, if the 1-forms $m^{A}$ are just $\fvierbein^A$ or $\thetavierbein^A$ (or any combination of these) then one has at linear order
\ba
m^{A}\wedge t_{A}\propto \text{tr}(\hat{e}_{(1)})\varepsilon=h^{A}{}_{A}\varepsilon ,
\ea
which yields $h^{A}{}_{A}=0$, as expected.

\section{Conclusions}
\label{S6}
In this paper we have analyzed a vielbein formulation of a family of ghost free massive gravity theories obtained from the de Rham-Gabadaze-Tolley metric theories. Summarizing here only our results for $D=4$ dimensions and a non dynamical metric $f_{\mu \nu}$ which is flat, this family is a three parameter set of theories characterized by an arbitrary combination of three mass terms (\ref{massvier}) each proportional to a constant $\beta_n$, $n=1,2,3$, in addition to a possibly non vanishing cosmological constant $\Lambda$ (which is also given by a similar term proportional to $\beta_0$). In the vierbein formulation, starting from an arbitrary vierbein (hence with a priori 16 free polarizations) we showed that the constraints associated with local Lorentz invariance (encoded in the symmetry of the energy momentum tensor) and those (vector constraints) associated with the Bianchi identities generally reduce the number of physical polarizations to 6. We then studied the three cases in which all but one among $\beta_1$, $\beta_2$ and $\beta_3$ vanish. In the first case (only $\beta_1 \neq 0$), we began by showing that the field equations impose the vierbein to be symmetric; and then, we showed how an extra scalar constraint (a scalar combination of the field equations only containing first derivative) can be obtained, analogous to the tracelessness of the graviton field in the standard (quadratic) Fierz-Pauli theory. This is obtained by taking into account the vector constraints as well as a suitable trace over the field equations. This extra constraint further reduces the number of propagating degrees of freedom in agreement with results obtained using the full Hamitonian analysis of the theory  \cite{Hassan:2011ea,Hassan:2011zd,Hassan:2011tf,Hassan:2011hr}. In the second case (only $\beta_2$ non vanishing) a similar constraint can be obtained, with however the following two important differences: firstly one has to impose the symmetry of the vierbein which is not garanteed anymore by the field equations, secondly one uses a different trace. In the last case (only $\beta_3$ non vanishing), we were able to show that, even though the field equations do impose the symmetry of the vierbein, the previously followed procedure to obtain a scalar constraint does not work. The same would hence be true for an arbitrary combination of the three mass terms as soon as $\beta_3$ does not vanish, however a scalar constraint can be obtained for such a combination and a vanishing $\beta_3$. Let us stress that is it  quite remarkable and non trivial that such a scalar constraint can be obtained at all in some fairly general cases. However, the puzzling difference between cases does not by itself invalidate the results of \cite{Hassan:2011ea,Hassan:2011zd,Hassan:2011tf,Hassan:2011hr} which are claimed to be valid for an arbitrary theory in the dRGT family (even though it might also open a way to reconciliate the contradictory claims which have been made in the literature about the Boulware Deser ghost). Indeed, it might just be that the extra constraint cannot be written in the most general case as a simple space-time scalar equation involving only first derivatives (which is a stronger requirement than the existence of an extra constraint on a subset of propagating degrees of freedom). This requires more work and could also be checked by a proper Hamitonian analysis of the vierbein formulation.

\begin{appendix}

\section{A simple way of extracting propagating degrees of freedom with vierbeins} 
\label{appC}
Our starting point here is the subset of theories with actions (in vacuum) given by $S_\kappa$ of Eq. (\ref{detxbis}). 
It is well known  that the Einstein Hilbert piece of this action can simply be rewritten as a term quadratic in the first derivatives of the vierbeins, such that the $S_{\kappa}$ action is
also given by 
\ba 
S_\kappa &=& M_P^2 \int d^4 x  \left[\thetavierbein\ G^{AB\mu \nu \lambda \rho} \thetavierbein_{A \mu,\nu} \thetavierbein_{B \lambda, \rho}
-  m^2 {\rm det} \left( \kappa \fvierbein^A_{\hphantom{A}\nu} - \thetavierbein^A_{\hphantom{A}\nu}\right)\right]
\ea 
where $G^{AB\mu \nu \lambda \rho}$ is a (mixed: space-time and Lorentz) tensor containing vierbeins $\thetavierbein^{A}$ (and their inverse) but not their derivatives. Hence, when one considers a {\it fixed} background vierbein $\fvierbein^{A}$, one can redefine the dynamical vierbein to be 
\ba
\tilde{\thetavierbein}^A{}_\mu = \thetavierbein^A{}_\mu - \kappa \fvierbein^A{}_\mu, 
\ea
so that the above action becomes
\ba \label{newact}
S_\kappa &=& M_P^2 \int d^4 x \left[\tilde{\thetavierbein}\ \tilde{G}^{AB\mu \nu \lambda \rho} \tilde{\thetavierbein}_{A \mu,\nu} \tilde{\thetavierbein}_{B \lambda, \rho}
-  m^2 {\rm det} \left(  \tilde{\thetavierbein}^A_{\hphantom{A}\nu}\right)\right] + S_{mix}
\ea 
where $S_{mix}$ involves a mixing between derivatives of $\tilde{\thetavierbein}^{A}$ and $\fvierbein^{A}$ (as well as a term containing only derivatives of $
\fvierbein^{A}$), and $\tilde{G}^{AB\mu \nu \lambda \rho}$ is obtained from $G^{AB\mu \nu \lambda \rho}$ by replacing $\thetavierbein^{A}$ with $\tilde{\thetavierbein}^{A}$. When $\fvierbein^A=dx^A$, $S_{mix}$ is simply given by 
\ba \label{Smixpart}
S_{mix} &=& M_P^2 \int d^4 x \tilde{\thetavierbein}  \tilde{K}^{AB\mu \nu \lambda \rho} \tilde{\thetavierbein}_{A \mu,\nu} \tilde{\thetavierbein}_{B \lambda, \rho}
\ea
where $\tilde{K}^{AB\mu \nu \lambda \rho}$ depends on $\thetavierbein^{A}$ and $\fvierbein^{A}$ but not on their derivatives. 
Notice now that the first two terms in (\ref{newact}) are invariant under diffeomorphisms (the mass term has been replaced by a cosmological constant), and that the term (\ref{Smixpart}) can just be interpreted as a modification to the kinetic term of the massless graviton, in particular it explicitly depends on the background vierbein $\fvierbein^{A}$. As such (together with the first term in (\ref{newact}) it encodes for the kinetic terms of the extra propagating polarization of the massive graviton.

\section{Field equations in the vielbein formulation}
\label{appA}
We vary the action (\ref{action}) with respect to the vielbein $\thetavierbein^{A}$. Note that for simplicity we have set $M_{P}=m^{2}=1$. This yields
\be
\delta S = \int \delta\Omega^{AB}\wedge\thetavierbein^{*}_{AB}+\int\Omega^{AB}\wedge\delta\thetavierbein^{*}_{AB}-\sum_{n=0}^{D-1}\beta_{n}\int \fvierbein^{A_{1}}\wedge\dots\wedge \fvierbein^{A_{n}}\wedge\delta\thetavierbein^{*}_{A_{1}\dots A_{n}}\ .
\ee
Using that $\delta\thetavierbein^{*}_{A_{1}\dots A_{n}}=(-1)^{n}\delta\thetavierbein^{A}\wedge\thetavierbein^{*}_{AA_{1}\dots A_{n}}$ the last term can be rewritten as
\be
\int\delta\thetavierbein^{A}\wedge\sum_{n=0}^{D-1}\beta_{n} \fvierbein^{A_{1}}\wedge\dots \fvierbein^{A_{n}}\wedge\thetavierbein^{*}_{AA_{1}\dots A_{n}}=2\int \delta\thetavierbein^{A}\wedge t_{A}\ .
\ee
The same method shows that the second term reads
\be
\int\delta\thetavierbein^{A}\wedge\Omega^{BC}\wedge\thetavierbein^{*}_{ABC}=2\int\delta\thetavierbein^{A}\wedge G_{A}\ .
\ee
Thus the only potential complications arise from the first term. Let us look at it more closely. Using the definition of the curvature 2-form as well as an integration by parts we can rewrite it  
\be
\int \delta\omega^{AB}\wedge d\thetavierbein^{*}_{AB}+\int\delta\omega^{A}{}_{C}\wedge\omega^{CB}\wedge\thetavierbein^{*}_{AB}+\int \omega^{A}{}_{C}\wedge\delta\omega^{CB}\wedge\thetavierbein^{*}_{AB}\ .
\label{cancel}
\ee 
Using the following straightforward consequence of the torsion-free condition
\be
\mathcal{D}\thetavierbein^*_{A_1\dots A_n}= d\thetavierbein^{*}_{A_1\dots A_n}-\sum_{k=1}^{n}\omega^{B}{}_{A_{k}}\wedge\thetavierbein^{*}_{A_{1}\dots A_{k-1}BA_{k}\dots A_{n}}=0,
\ee
we see that
\be
d\thetavierbein^{*}_{AB}=d\thetavierbein^{C}\wedge\thetavierbein^{*}_{ABC}=-\omega^{C}{}_{D}\wedge\thetavierbein^{D}\wedge\thetavierbein^{*}_{ABC}=\omega^{C}{}_{A}\wedge\thetavierbein^{*}_{CB}+\omega^{C}{}_{B}\wedge\thetavierbein^{*}_{AC}
\ee 
and the expression (\ref{cancel}) becomes
\be
\begin{split}
&\int \delta\omega^{AB}\wedge\omega^{C}{}_{A}\wedge\thetavierbein^{*}_{CB}+\int \delta\omega^{AB}\wedge\omega^{C}{}_{B}\wedge\thetavierbein^{*}_{AC}\\
&+\int\delta\omega^{A}{}_{C}\wedge\omega^{CB}\wedge\thetavierbein^{*}_{AB}+\int \omega^{A}{}_{C}\wedge\delta\omega^{CB}\wedge\thetavierbein^{*}_{AB}\ ,
\end{split}
\ee
which is clearly equal to zero. Therefore we have proven that
\be
\delta S =2\int\delta\thetavierbein^{A}\wedge (G_{A}-t_{A})\ ,
\ee
which naturally implies the field equations (\ref{eom}).

\section{On the case $\beta_3 \neq 0$} 
 \label{appB}
We show here that when $\beta_{3}$ is the only non-vanishing $\beta$ (cf. subsection \ref{b3}) there is no family of one forms $m^{A}$ which are polynomial in 
$\thetavierbein^A$ and $\fvierbein^A$ such that $m^{A}\wedge G_{A}=m^{A}\wedge t_{A}$ provides an additional constraint in the sense discussed above, i.e. such that $m^{A}\wedge G_{A}$ does not contain second order derivatives. Using $m^{A}\equiv m^{A}{}_{B}\thetavierbein^{B}$ and the $\sim$ notation introduced earlier we have, for an arbitrary collection of one forms $m^A$
\be
\begin{split}
m^{A}\wedge G_{A}&=\frac{1}{2} m^{A}\wedge\Omega^{BC}\wedge\thetavierbein^{*}_{ABC}\\
&=\frac{1}{2}m^{A}{}_{D}\W^{BC}{}_{EF}\thetavierbein^{D}\wedge\thetavierbein^{E}\wedge\thetavierbein^{F}\wedge\thetavierbein^{*}_{ABC}\\
&=\frac{1}{2}m^{A}{}_{D}\W^{BC}{}_{EF}\delta^{DEF}_{ABC}\varepsilon\\
&\sim m^{A}{}_{D}\ e_{[E}{}^\mu \partial_\mu(w^{BC}{}_{F]})\delta^{DEF}_{ABC}\varepsilon\\
&\sim e_{D}{}^\mu \partial_\mu(m^{A}{}_{E}w^{BC}{}_{F})\delta^{DEF}_{ABC}\varepsilon\\
&\sim e_{A}{}^\mu\partial_\mu(m^{A}{}_{B}w^{BC}{}_{C}+m^{B}{}_{C}w^{AC}{}_{B}-m^{C}{}_{C}w^{AB}{}_{B})\varepsilon\ .
\end{split}
\ee
We show then that it is not possible to find $m^{A}$ (provided it is only built from $\fvierbein^{A}$ and $\thetavierbein^{A}$) such that the term in factor of $\beta_3 \varepsilon$ in Eq.(\ref{n=3}) can be written 
as the term in the parentheses above. More specifically, we prove that there do not exist invertible matrices $\mathcal{M}=(m^{A}{}_{B})_{1\leq A,B\leq d}$ and $\mathcal{S}=(s^{A}{}_{B})_{1\leq A,B\leq d}$ such that for all $\thetavierbein^{A}{}_{B}$
\be
s^{A}{}_{C}w^{BC}{}_{D}\thetavierbein^{D}{}_{B}=m^{A}{}_{B}w^{BC}{}_{C}+m^{B}{}_{C}w^{AC}{}_{B}-m^{C}{}_{C}w^{AB}{}_{B}.
\ee 
Indeed, let us first rewrite the above equation in the more convenient way
\be
(s^{A}{}_{C}\thetavierbein^{D}{}_{B}-m^{A}{}_{B}\delta^{D}{}_{C}+m^{D}{}_{B}\delta^{A}{}_{C}-m^{E}{}_{E}\delta^{A}{}_{C}\delta^{D}{}_{B})w^{BC}{}_{D}=0 ,
\ee
and notice that we can treat $w^{BC}{}_{D}$ as an indeterminate (this because one can vary the derivatives of the $\thetavierbein^{A}$ without varying the $\thetavierbein^{A}$ or the $e_{A}$). The first subtlety that we have to take into account is that $w^{BC}{}_{D}$ is antisymmetric in its first two indices. A second subtlety arises as a consequence of the symmetry of $e^{AB}$ or equivalently of its inverse. Indeed in this case $w^{BC}{}_{D}$ acquires a new symmetry which reads $w_{[B|CD|}\thetavierbein^{C}{}_{E}\thetavierbein^{D}{}_{F]}=0$ (this relation is just a consequence of the differentiation of $\eta_{AB}\thetavierbein^{A}\wedge dx^{B}=0$). Therefore the identification reads
\be
s^{A}{}_{[C}\thetavierbein^{D}{}_{B]}-m^{A}{}_{[B}\delta^{D}{}_{C]}+m^{D}{}_{[B}\delta^{A}{}_{C]}-m^{E}{}_{E}\delta^{A}{}_{[C}\delta^{D}{}_{B]}=\eta_{G[B}\eta_{C]H}\Lambda^{A}{}_{K}\epsilon^{KEFG}\thetavierbein^{H}{}_{E}\thetavierbein^{D}{}_{F}\ ,
\ee
where $\Lambda^{A}{}_{B}$ is an arbitrary rank two tensor which is a function of the $E^{A}{}_{B}$.
Taking two traces over the indices $A$, $B$ and $C$, $D$ simultaneously and introducing the matrix $\mathcal{E}\equiv (\thetavierbein^{A}{}_{B})_{1\leq A,B\leq d}$ we get
\be
\text{tr}(\mathcal{M})=\frac{1}{6}(\text{tr}(\mathcal{S})\text{tr}(\mathcal{E})-\text{tr}(\mathcal{E} \mathcal{S}))\ .
\ee
Now, taking only one trace first over the indices $A$, $B$ and then over $C$, $D$, and plugging in the above result we get
\be
\begin{split}
\mathcal{M}&=\frac{1}{2}(\mathcal{E} \mathcal{S}-\mathcal{E}\text{tr}(\mathcal{S}))+\frac{1}{6}(\text{tr}(\mathcal{S})\text{tr}(\mathcal{E})-\text{tr}(\mathcal{S}\mathcal{E}))I-\frac{1}{2}\mathcal{X}\\
&=\frac{1}{2}(\mathcal{S}\mathcal{E}-\mathcal{S}\text{tr}(\mathcal{E}))+\frac{1}{6}(\text{tr}(\mathcal{S})\text{tr}(\mathcal{E})-\text{tr}(\mathcal{S}\mathcal{E}))I\ ,
\end{split}
\label{2TRACES}
\ee
where $\mathcal{X}$ is the matrix whose matrix elements are the 
\be
\eta_{G[A}\eta_{C]H}\Lambda^{A}{}_{K}\epsilon^{KEFG}\thetavierbein^{H}{}_{E}\thetavierbein^{D}{}_{F}=\Lambda_{GK}\epsilon^{KEFG}\thetavierbein_{CE}\thetavierbein^{D}{}_{F}-\eta_{GC}\Lambda_{HK}\epsilon^{KEFG}\thetavierbein^{H}{}_{E}\thetavierbein^{D}{}_{F}\ .
\label{PIRATE}
\ee
But since the $\Lambda^{A}{}_{B}$ are functions of the $E^{A}{}_{B}$ then necessarily the terms $\Lambda_{GK}$ and $\thetavierbein^{H}{}_{E}\Lambda_{HK}$ must be symmetric in their free indices. Thus the right hand side of Eq. (\ref{PIRATE}) must be identically zero which shows that we may disregard the additional $\mathcal{X}$ term in (\ref{2TRACES}) and obtain 
\be
\mathcal{S}\mathcal{E} -\mathcal{S}\text{tr}(\mathcal{E})=\mathcal{E} \mathcal{S}-\mathcal{E}\text{tr}(\mathcal{S})\ .
\ee 
The only $\mathcal{S}$ which verifies this relation whatever the $\mathcal{E}$ is $\mathcal{S}=\alpha\mathcal{E}$. This gives us an associated matrix $\mathcal{M}$ and it suffices to verify that this combination does not work in order to complete the proof.

\end{appendix}


\begin{thebibliography}{99}


\bibitem{Dvali:2000hr} 
  G.~R.~Dvali, G.~Gabadadze and M.~Porrati,
  Phys.\ Lett.\ B {\bf 485}, 208 (2000)
  [hep-th/0005016].
  
\bibitem{Deffayet:2000uy} 
  C.~Deffayet,
  Phys.\ Lett.\ B {\bf 502}, 199 (2001)
  [hep-th/0010186].

\bibitem{Deffayet:2001pu} 
  C.~Deffayet, G.~R.~Dvali and G.~Gabadadze,
  Phys.\ Rev.\ D {\bf 65}, 044023 (2002)
  [astro-ph/0105068].

\bibitem{Rubakov:2008nh} 
  V.~A.~Rubakov and P.~G.~Tinyakov,
  Phys.\ Usp.\  {\bf 51}, 759 (2008)
  [arXiv:0802.4379 [hep-th]].


\bibitem{Hinterbichler:2011tt} 
  K.~Hinterbichler,
  arXiv:1105.3735 [hep-th].



\bibitem{Fierz:1939ix}M. Fierz, Helv. Phys. Acta 12 (1939) 3;
M.~Fierz and W.~Pauli,
Proc.\ Roy.\ Soc.\ Lond.\ A {\bf 173}, 211 (1939).

\bibitem{vanDam:1970vg}
H.~van Dam and M.~J.~G.~Veltman,
Nucl.\ Phys.\ B {\bf 22}, 397 (1970).
V.~I.~Zakharov, JETP Lett. {\bf 12}, 312 (1970)~.
 Y.~Iwasaki,
Phys.\ Rev.\ D {\bf 2} (1970) 2255.

\bibitem{Vainshtein:1972sx}
A.~I.~Vainshtein,
Phys.\ Lett.\ B {\bf 39} (1972) 393.


\bibitem{Deffayet:2001uk}
  C.~Deffayet, G.~R.~Dvali, G.~Gabadadze and A.~I.~Vainshtein,
  Phys.\ Rev.\ D {\bf 65} (2002) 044026
  [hep-th/0106001].

\bibitem{Babichev:2009us} 
  E.~Babichev, C.~Deffayet and R.~Ziour,
  JHEP {\bf 0905}, 098 (2009)
  [arXiv:0901.0393 [hep-th]].
  
\bibitem{Babichev:2009jt} 
  E.~Babichev, C.~Deffayet and R.~Ziour,
  Phys.\ Rev.\ Lett.\  {\bf 103}, 201102 (2009)
  [arXiv:0907.4103 [gr-qc]].

\bibitem{Babichev:2010jd} 
  E.~Babichev, C.~Deffayet and R.~Ziour,
  Phys.\ Rev.\ D {\bf 82}, 104008 (2010)
  [arXiv:1007.4506 [gr-qc]].


\bibitem{Boulware:1973my} 
  D.~G.~Boulware and S.~Deser,
  Phys.\ Rev.\ D {\bf 6}, 3368 (1972).

\bibitem{Creminelli:2005qk} 
  P.~Creminelli, A.~Nicolis, M.~Papucci and E.~Trincherini,
  JHEP {\bf 0509}, 003 (2005)
  [hep-th/0505147].

\bibitem{deRham:2010kj} 
  C.~de Rham, G.~Gabadadze and A.~J.~Tolley,
  Phys.\ Rev.\ Lett.\  {\bf 106}, 231101 (2011)
  [arXiv:1011.1232 [hep-th]].

\bibitem{deRham:2010ik} 
  C.~de Rham and G.~Gabadadze,
  Phys.\ Rev.\ D {\bf 82}, 044020 (2010)
  [arXiv:1007.0443 [hep-th]].

\bibitem{deRham:2011rn} 
  C.~de Rham, G.~Gabadadze and A.~Tolley,
  arXiv:1107.3820 [hep-th].
  
\bibitem{ArkaniHamed:2002sp} 
  N.~Arkani-Hamed, H.~Georgi and M.~D.~Schwartz,
  Annals Phys.\  {\bf 305}, 96 (2003)
  [hep-th/0210184].
  
\bibitem{Deffayet:2005ys} 
  C.~Deffayet and J.~-W.~Rombouts,
  Phys.\ Rev.\ D {\bf 72}, 044003 (2005)
  [gr-qc/0505134].
  


\bibitem{Alberte:2010qb}
  L.~Alberte, A.~H.~Chamseddine and V.~Mukhanov,
  JHEP {\bf 1104} (2011) 004
  [arXiv:1011.0183 [hep-th]].

\bibitem{Chamseddine:2011mu} 
  A.~H.~Chamseddine and V.~Mukhanov,
  JHEP {\bf 1108}, 091 (2011)
  [arXiv:1106.5868 [hep-th]].


\bibitem{Hassan:2011ea}
  S.~F.~Hassan and R.~A.~Rosen,
  arXiv:1111.2070 [hep-th].

\bibitem{Hassan:2011zd}
  S.~F.~Hassan and R.~A.~Rosen,
  JHEP {\bf 1202} (2012) 126
  [arXiv:1109.3515 [hep-th]].

\bibitem{Hassan:2011tf}
  S.~F.~Hassan, R.~A.~Rosen and A.~Schmidt-May,
  JHEP {\bf 1202} (2012) 026
  [arXiv:1109.3230 [hep-th]].

\bibitem{Hassan:2011hr}
  S.~F.~Hassan and R.~A.~Rosen,
  Phys.\ Rev.\ Lett.\  {\bf 108} (2012) 041101
  [arXiv:1106.3344 [hep-th]].

\bibitem{Kluson} 
  J.~Kluson,
  JHEP {\bf 1201}, 013 (2012)
  [arXiv:1109.3052 [hep-th]].
   J.~Kluson,
  arXiv:1112.5267 [hep-th].
   J.~Kluson,
  arXiv:1202.5899 [hep-th].
  J.~Kluson,
  arXiv:1204.2957 [hep-th].
  
  \bibitem{OthersCounting}
  D.~Comelli, M.~Crisostomi, F.~Nesti and L.~Pilo,
  arXiv:1204.1027 [hep-th].
  S.~F.~Hassan, A.~Schmidt-May and M.~von Strauss,
  arXiv:1203.5283 [hep-th].
  A.~Golovnev,
  Phys.\ Lett.\ B {\bf 707}, 404 (2012)
  [arXiv:1112.2134 [gr-qc]].
  M.~Mirbabayi,
  arXiv:1112.1435 [hep-th].
  
\bibitem{Hinterbichler:2012cn}
  K.~Hinterbichler and R.~A.~Rosen,
  arXiv:1203.5783 [hep-th].
  
\bibitem{Hassan:2012wt}
  S.~F.~Hassan, A.~Schmidt-May and M.~von Strauss,
  arXiv:1204.5202 [hep-th].
  
  
\bibitem{Nibbelink:2006sz} 
  S.~Nibbelink Groot, M.~Peloso and M.~Sexton,
  Eur.\ Phys.\ J.\ C {\bf 51}, 741 (2007)
  [hep-th/0610169].
  
  
  \bibitem{Damour:2002ws}
T.~Damour and I.~I.~Kogan,
Phys.\ Rev.\ D {\bf 66} (2002) 104024
[arXiv:hep-th/0206042].

\bibitem{Hassan:2011vm} 
  S.~F.~Hassan and R.~A.~Rosen,
  JHEP {\bf 1107}, 009 (2011)
  [arXiv:1103.6055 [hep-th]].


\bibitem{sqrt}
N.~J.~Higham, 
Linear Algebra and its Applications, 88-89, (1987).
  J.~Gallier, 
  arXiv:0805.0245 [math-GM].

\bibitem{usbis}
 C.~Deffayet, J.~Mourad and G.~Zahariade,
  arXiv:1208.4493 [gr-qc].

\bibitem{DuboisViolette:1986ws} 
  M.~Dubois-Violette and J.~Madore,
  Commun.\ Math.\ Phys.\  {\bf 108}, 213 (1987).

\bibitem{Volkov:2012wp}
  M.~S.~Volkov,
  arXiv:1202.6682 [hep-th].



\end{thebibliography}
\end{document}